\documentclass[letterpaper, 10 pt, conference]{ieeeconf}  % Comment this line out if you need a4paper

\IEEEoverridecommandlockouts                              % This command is only needed if 
\usepackage{amsmath} % assumes amsmath package installed
\usepackage[hidelinks]{hyperref}
\usepackage{graphicx}
\usepackage{subfigure}
\usepackage{cite}
\usepackage{todonotes}
\title{\LARGE \bf Homography-based Visual Servoing with Remote Center of Motion for Semi-autonomous Robotic Endoscope Manipulation}

\author{Martin Huber$^{1}$, John Bason Mitchell$^{1,2}$, Ross Henry$^{1}$, S\'{e}bastien Ourselin$^{1}$,\\Tom Vercauteren$^{1}$, and Christos Bergeles$^{1}$% <-this % stops a space
\thanks{$^{1}$School of Biomedical Engineering \& Image Sciences, Faculty of Life Sciences \& Medicine,
        King's College London, London, United Kingdom
        {\tt\small martin.huber@kcl.ac.uk}}%
\thanks{$^{2}$Department of Medical Physics and Biomedical Engineering, Faculty of Engineering Sciences, University College London, London, United Kingdom}
}

\DeclareMathOperator{\diag}{diag}

\begin{document}
%Watermark
%\SetWatermarkText{\textsf{\textbf{\wmtxt}}}

\maketitle
\thispagestyle{empty}
\pagestyle{empty}

%%%%%%%%%%%%%%%%%%%%%%%%%%%%%%%%%%%%%%%%%%%%%%%%%%%%%%%%%%%%%%%%%%%%%%%%%%%%%%%%
\begin{abstract}
% The dominant endoscope manipulation automation approaches in Minimally Invasive Surgery (MIS) rely on point positions wrt. the camera frame to infer a control policy
The dominant visual servoing approaches in Minimally Invasive Surgery (MIS) follow single points or adapt the endoscope's field of view based on the surgical tools' distance. These methods rely on point positions with respect to the camera frame to infer a control policy. Deviating from the dominant methods, we formulate a robotic controller that allows for image-based visual servoing that requires neither explicit tool and camera positions nor any explicit image depth information. The proposed method relies on homography-based image registration, which changes the automation paradigm from point-centric towards surgical-scene-centric approach. It simultaneously respects a programmable Remote Center of Motion (RCM). Our approach allows a surgeon to build a graph of desired views, from which, once built, views can be manually selected and automatically servoed to irrespective of robot-patient frame transformation changes. We evaluate our method on an abdominal phantom and provide an open source ROS Moveit integration for use with any serial manipulator\footnote[3]{ \url{https://github.com/RViMLab/h_rcm_vs_ws.git}}. A video is provided\footnote[4]{\label{foot:vid}\url{https://drive.google.com/file/d/1UCr__R2_7xit6TTq3T9pTIEg1fMsfT3j/view?usp=sharing}}.
\end{abstract}

%%%%%%%%%%%%%%%%%%%%%%%%%%%%%%%%%%%%%%%%%%%%%%%%%%%%%%%%%%%%%%%%%%%%%%%%%%%%%%%%
%\section{NOTES}
%\begin{itemize}
%    \item Video: 15th of August
%    \item Submission \url{http://www.ismr.gatech.edu/submissions}
%\end{itemize}

\section{INTRODUCTION}
When compared to open surgery, MIS takes place under endoscopic guidance and offers improved cosmetics, less blood loss, shorter recovery times and reduced cost \cite{vitiello2012emerging}. In a traditional MIS setup, the surgeon is supported by an assistant who guides the endoscope. Although this task is conceptually simple, it requires trained personnel, which introduces cost \cite{horgan2001robots}. The assistant surgeon exhibits tremor, suffers fatigue, and can be prone to communication failures \cite{horgan2001robots, palep2009robotic, li2020accelerated}. 

Several robotic endoscope holders, such as AESOP \cite{unger1994aesop}, ViKY \cite{long2007development}, and EndoAssist \cite{gilbert2009endoassist}, have been developed to address these shortcomings. Research in \cite{aiono2002controlled} and \cite{voros2010viky} showcased a reduction in the intervention time. While robotic endoscope holders can facilitate improvements, they introduce additional workload to the surgeon. With the advance of automated surgical systems this additional workload can be reduced \cite{moustris2011evolution}. Therefore, different methods to automate endoscopic camera motion were explored. 

Alongside automation via kinematic data, visual servoing, i.e. control through images, is considered a promising alternative, as it provides intra-operative feedback \cite{pandya2014review} and is less prone to errors from model mismatch \cite{azizian2014visual}. In semi-autonomous setups, such as gaze or voice control \cite{taniguchi2010classification}, visual servoing can robustly reflect a surgeon's intent and respect anatomical constraints or facilitate full autonomy.

Visual servoing approaches that satisfy a RCM constraint can be split into methods that rely on a mechanical RCM and methods that rely on a programmable RCM. There has been less research on visual servoing with programmable RCM because of robot singularities and constraints on the robot positioning, however, in contrast to a mechanical RCM, a programmable RCM can be adapted in real-time, and the robot, with which the programmable RCM is achieved, can be used for multiple purposes \cite{kuo2012kinematic}, for example in open surgery. Existing methods with mechanical RCM, and programmable RCM, will be detailed in Sec.\ \ref{sec:mech_rcm}, and Sec.\ \ref{sec:prog_rcm}, respectively.
\begin{figure}[t]
    \centering
    \includegraphics[scale=0.33]{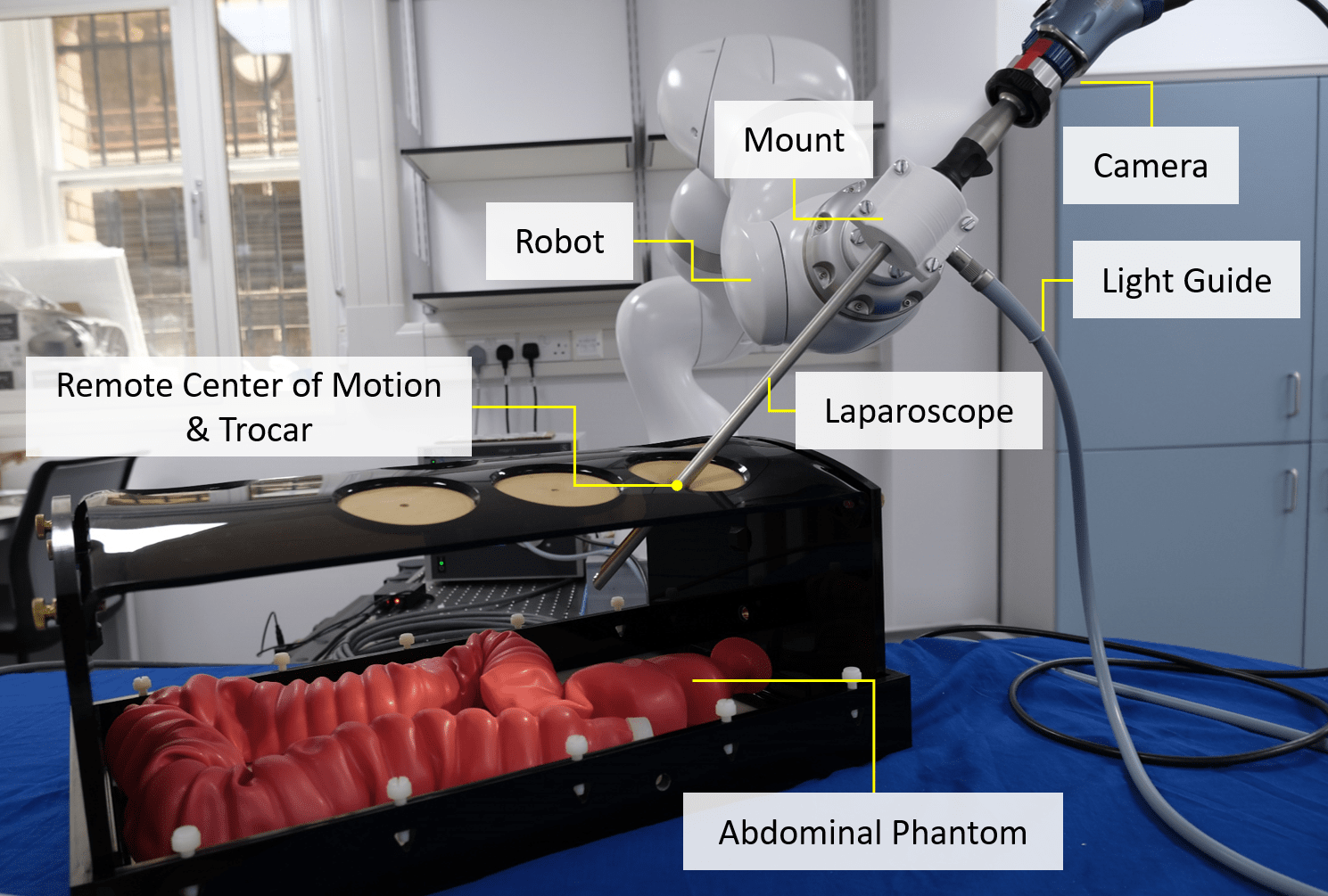}
    \caption{Robotic setup. A Storz Endocameleon Hopkins Telescope, which provides a light source port and a camera attachment point, is mounted to a KUKA LBR Med 7 R800 robot via a 3D printed clamp. The robotic system undergoes image-based control to reach desired views of the surgical scene and simultaneously pivots around a programmable RCM.}
    \label{fig:setup}
\end{figure}

%\subsection{with Mechanical RCM}
\subsection{Visual Servoing with Mechanical RCM}
\label{sec:mech_rcm}
Examples of approaches that use a mechanical RCM are~\cite{omote1999self}, where a visual servo controls the position of a marked forceps in image space. In \cite{agustinos2014visual, voros2007automatic}, the tool entry point is exploited to find the tool tip in image space and to center it via visual servoing. Another common scheme is to alter the camera's zoom based on the surgical tools' distance, which was first presented in \cite{king2013towards}, where the tools are tracked with markers. Research in \cite{eslamian2020development, mariani2020experimental, dascan}, based on  \cite{Eslamian2016TowardsTI, eslamian2017autonomous}, adjusts the camera's distance in this manner. They align the camera's optical axis with the line that spans from RCM to the tools' center point. Such an approach requires a complicated registration procedure. In~\cite{abdelaal2020orientation}, Abdelaal \emph{et al.} also adjust the camera's distance to the surgical scene based on the tool distance, but they align the camera's optical axis with the scene's surface normal, which is made possible by their 6 DOF endoscope. Yu \emph{et al.}~\cite{yu2016automatic} adjust the field of view's width based on tool distance. In~\cite{ma2019autonomous}, Ma \emph{et al.} deploy a visual servo to center a marked tool by incorporating depth information, which they extract from camera and tool motion. In \cite{ma2020visual}, they extend this work into a quadratic program in which they constrain the camera's distance with respect to the tools and the tool position in the image plane, whilst minimizing the joint velocities. They rely on stereoscopic images for depth information.

\subsection{Visual Servoing with Programmable RCM}
\label{sec:prog_rcm}
Multi-purpose serial manipulators can achieve a RCM programmatically. In \cite{osa2010framework}, Osa \emph{et al.} adapt the interaction matrix to account for the RCM constraint, which they then use to control a point in image space. The authors in \cite{aghakhani2013task} design a composite Jacobian method that integrates a RCM objective with a task function that defines an error on points in image space. Yang \emph{et al.} in \cite{yang2019adaptive} also design a Jacobian gain controller that enforces the tip of a tool to reside within a defined region. They additionally request the endoscope to extend the surgeon's natural line of sight. In \cite{li2020accelerated}, Li \emph{et al.} introduce the RCM and a visual error via the image Jacobian as constraints to a quadratic problem that aims at satisfying these constraints whilst minimizing the joint velocities.

% option A:
% Automation in General
% Automation with mechanical RCM, Automation with programmable RCM
% Add other kinematic papers (\cite{weede2011intelligent})
% Add sun
% Move kinematic shortcomings to "Limitations of current approaches"

% option B:
% Remove "kinematic" VS from Mech RCM
% Add short list of kinematic approaches
% Add collection of visual servo with mechanical rcm
% Add sun
% Highlight shortcomings

%\cite{weede2011intelligent}   % Markov
%\cite{sun2020development}     % kinematic -> image space velocity, center tool
%\cite{sun2020adaptive}        % mechanical rcm, adjust angle based on fusion of kinematic and visual data (stereoscopic)
%\cite{sun2020visual}          % (estimate depth based on laparoscopic motion)

\subsection{Limitations of Current Approaches and Contributions}
The majority of existing methods rely on the tool distance to infer a control law. Only in \cite{ma2019autonomous, ma2020visual, aghakhani2013task, yang2019adaptive, li2020accelerated, osa2010framework}, the position of arbitrary points wrt. the camera frame is fed back to the robot. All of the existing methods rely on relative positions, which either requires tool and camera positions or depth images. Position data might only be accessible in a fully robotic setup and image depth is difficult to estimate in a dynamic surgical environment from a monocular camera. Stereoscopic images are usually not available in robotically assisted surgery.

Our paper addresses the above limitations with the following contributions:
\begin{itemize}
    \item We introduce a visual servo that navigates towards desired images rather than towards points.
    \item We formulate a visual servo control law that depends neither on explicit tool and camera positions nor on depth information.
\end{itemize} 
These are achieved with a programmable RCM, as it, in contrast to a mechanical RCM, is more flexible. 

This paper is structured as follows. In Sec.\ \ref{sec:methods}, we introduce the necessary theoretical background and the derivation of the proposed visual servoing task. In Sec.\ \ref{sec:experimental_setup}, we explain implementation details and the robotic setup. Results are provided in Sec.\ \ref{sec:results}, and conclusions in Sec.\ \ref{sec:conclusions}.

\section{METHODS}
\label{sec:methods}
%In this section,
Here, we first introduce the composite Jacobian for control in Sec.\ \ref{sec:task_rcm}. Then, we extend it by a novel homography-based task function in Sec.\ \ref{sec:homography_task}, and describe the processing pipeline in Sec.\ \ref{sec:pipe}. In the following, scalars are depicted by lower case letters, vectors through bold lower case letters, and matrices as bold upper case letters. A point $x$ is described with respect to frame F as $^\text{F}\mathbf{x}$.

\subsection{Task Control with Remote Center of Motion Objective}
\label{sec:task_rcm}
For the task control with RCM objective, we follow the derivation of Aghakhani \emph{et al.}  \cite{aghakhani2013task}. Therefore, as schematically shown in Fig.\ \ref{fig:schematic}, an open kinematic chain is attached to reference frame W. An endoscope is attached to the chain. It originates at position $^\text{W}\mathbf{x}_i$ and has its camera frame at position $^\text{W}\mathbf{x}_{i+1}$. The endoscope enters the patient through the trocar at position $^\text{W}\mathbf{x}_\text{trocar}$. The RCM position $^\text{W}\mathbf{x}_\text{RCM}$ is required to lie along the line connecting $^\text{W}\mathbf{x}_i$ to $^\text{W}\mathbf{x}_{i+1}$, hence
\begin{equation}
^\text{W}\mathbf{x}_\text{RCM} = ^\text{W}\mathbf{x}_i+\lambda\left(^\text{W}\mathbf{x}_{i+1} - ^\text{W}\mathbf{x}_i\right),
\label{eq:lambda}
\end{equation}
where the scalar $\lambda \geq 0$ is proportional to the entry depth. $\lambda = 0$ corresponds to maximal insertion. The endoscope's translational velocity at position $^\text{W}\mathbf{x}_\text{RCM}$ has to remain zero for the endoscope to reside at the trocar $^\text{W}\mathbf{x}_\text{trocar}$. It was derived in \cite{aghakhani2013task} as
\begin{equation}
    ^\text{W}\dot{\mathbf{x}}_\text{RCM} = \begin{bmatrix}\mathbf{J}^v_i + \lambda(\mathbf{J}^v_{i+1}-\mathbf{J}^v_i)\\ ^\text{W}\mathbf{x}_{i+1} - ^\text{W}\mathbf{x}_i\end{bmatrix}^\text{T}\begin{bmatrix}\dot{\mathbf{q}} \\ \dot{\lambda}\end{bmatrix},
    \label{eq:dx_RCM}
\end{equation}
where $\mathbf{J}^v_i$, $\mathbf{J}^v_{i+1}$ are the Jacobians' top three rows, therefore the translational parts, corresponding to points $^\text{W}\mathbf{x}_i$, $^\text{W}\mathbf{x}_{i+1}$ w.r.t. the world frame, $\dot{\mathbf{q}}$ are the instantaneous joint velocities, and $\dot{\lambda}$ is the rate of change of entry depth. Eq.\ (\ref{eq:dx_RCM}) can be rewritten as
\begin{equation}
    ^\text{W}\dot{\mathbf{x}}_\text{RCM} = \mathbf{J}_\text{RCM}\begin{bmatrix}\dot{\mathbf{q}} \\ \dot{\lambda}\end{bmatrix}.
    \label{eq:dx_RCM_short}
\end{equation}
Expanding on \cite{aghakhani2013task}, we introduce a feedback to $\lambda$ by projecting the trocar position $\mathbf{x}_\text{trocar}$ onto the endoscope via
\begin{equation}
    \lambda = \frac{(^\text{W}\mathbf{x}_{i+1} - ^\text{W}\mathbf{x}_i)^\text{T}(^\text{W}\mathbf{x}_\text{trocar}-^\text{W}\mathbf{x}_i)}{||^\text{W}\mathbf{x}_{i+1}-^\text{W}\mathbf{x}_i||_2^2}.
\end{equation}
Eq.\ (\ref{eq:dx_RCM_short}) can be further extended by a task as follows
\begin{equation}
    \begin{bmatrix}\dot{\mathbf{t}} \\ ^\text{W}\dot{\mathbf{x}}_\text{RCM}\end{bmatrix} =
    \begin{bmatrix}\mathbf{J}_\text{t} & \mathbf{0}_{n_\text{t}\times 1} \\ \multicolumn{2}{c}{\mathbf{J}_\text{RCM}}
    \end{bmatrix}
    \begin{bmatrix}\dot{\mathbf{q}}\\\dot{\lambda}\end{bmatrix},
    \label{eq:task_jac}
\end{equation}
where $\dot{\mathbf{t}}$ is the task velocity with task dimension $n_\text{t}$ and $\mathbf{J}_\text{t}$ is the task Jacobian. Eq.\ (\ref{eq:task_jac}) can be turned into a PID controller
\begin{equation}
    \centering
    \begin{bmatrix}
        \dot{\mathbf{q}}\\
        \dot{\lambda}
    \end{bmatrix} = 
    \mathbf{J}_\text{cp}^\#
    \left(
        \mathbf{K}^\text{p}
        \begin{bmatrix}
            \mathbf{e}_\text{t}^\text{p}\\
            ^\text{W}\mathbf{e}_\text{RCM}^\text{p}
        \end{bmatrix} +
        \mathbf{K}^\text{i}
        \begin{bmatrix}
            \mathbf{e}_\text{t}^\text{i}\\
            ^\text{W}\mathbf{e}_\text{RCM}^\text{i}
        \end{bmatrix} +
        \mathbf{K}^\text{d}
        \begin{bmatrix}
            \mathbf{e}_\text{t}^\text{d}\\
            ^\text{W}\mathbf{e}_\text{RCM}^\text{d}
        \end{bmatrix}
    \right),
    \label{eq:pid}
\end{equation}
where $\mathbf{J}_\text{cp}^\#$ is the pseudo-inverse of the composite Jacobian from (\ref{eq:task_jac}), $\mathbf{e}^{\text{p}/\text{i}/\text{d}}_t$ and $^\text{W}\mathbf{e}^{\text{p}/\text{i}/\text{d}}_\text{RCM}$, are the proportional, integral, and differential errors for the task and the RCM, respectively, and $\mathbf{K}^{\text{p}/\text{i}/\text{d}}$ are the diagonal gain matrices. Therein, $^\text{W}\mathbf{e}_\text{RCM}^{\text{i}/\text{d}}$ are computed as the integral, and the differential of the proportional error $^\text{W}\mathbf{e}_\text{RCM}^\text{p} = ^\text{W}\mathbf{x}_\text{trocar} - ^\text{W}\mathbf{x}_\text{RCM}$. 

In the following section, we introduce a homography-based visual servoing task.

\subsection{Homography-based Visual Servoing Task}
\label{sec:homography_task}

\begin{figure}
    \centering
    \includegraphics[scale=0.3]{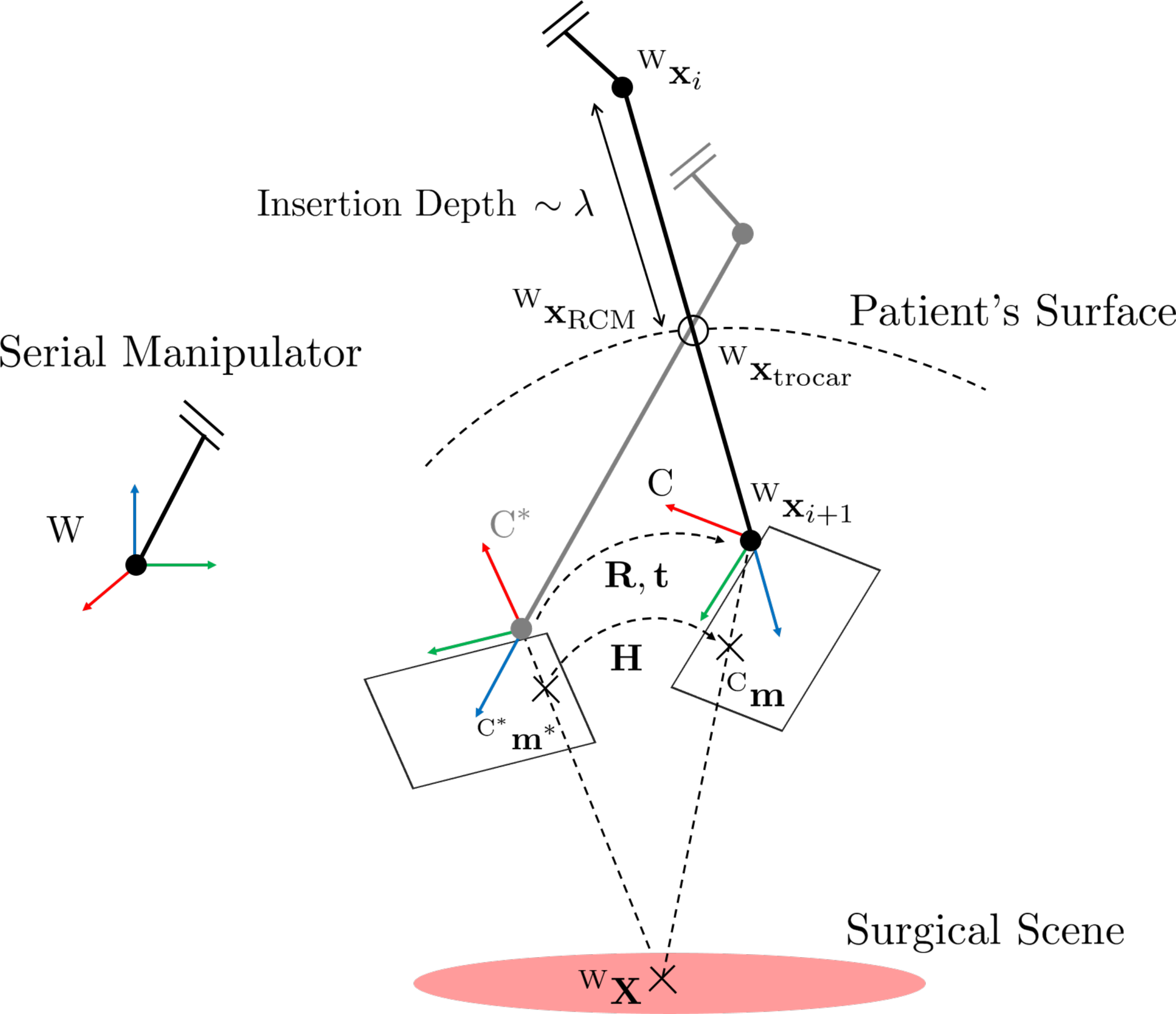}
    \caption{Schematic illustration of the setup: The axes' RGB coloring corresponds to XYZ, respectively. A serial manipulator is connected to the world frame W. The endoscope spans from $^\text{W}\mathbf{x}_i$ to $^\text{W}\mathbf{x}_{i+1}$ and it enters the trocar, which lies at $\mathbf{x}_\text{trocar}$. The camera rotates around the RCM $^\text{W}\mathbf{x}_\text{RCM}$ and its entry depth is proportional to $\lambda \geq 0$.  The camera observes the surgical scene (pink) from different frames $\text{C}$ and $\text{C}^*$.}
    \label{fig:schematic}
\end{figure}

\begin{figure*}[ht]
    \centering
    \includegraphics[scale=0.5]{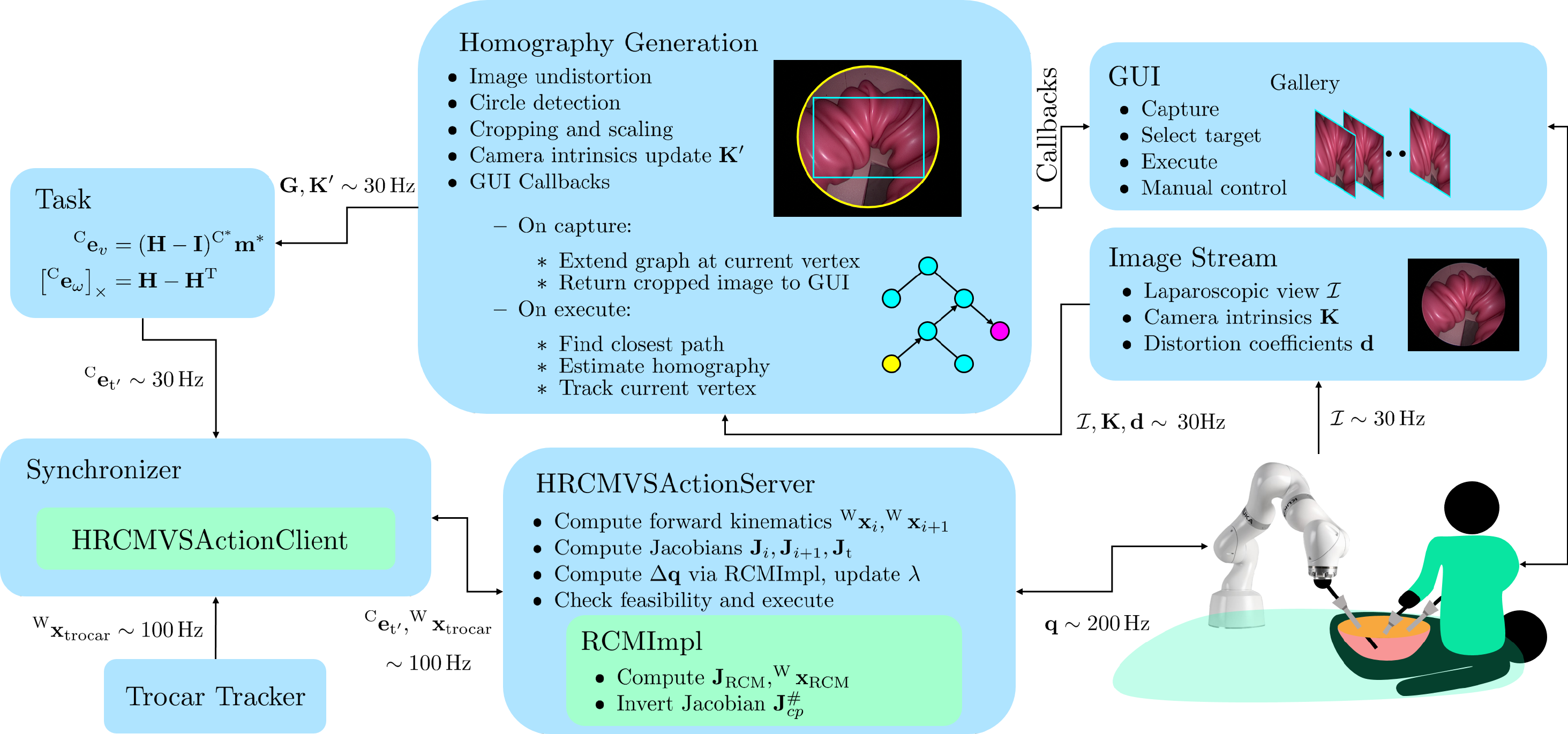}
    \caption{Processing pipeline. A surgeon manually controls the robot through a GUI, collecting desired views along the way. The images are pre-processed, and a graph of desired views is built in the background by the homography generation node. Once built, the surgeon selects desired views through the GUI, which triggers a shortest path finding from the current vertex (yellow), to the desired one (pink), and the execution of subsequent homography estimations that lead to the target.}
    \label{fig:pipe}
\end{figure*}

Suppose point $^\text{W}\mathbf{X}$ is projected from a plane, i.e. the surgical scene, onto normalized coordinates $\mathbf{m}^*$ in camera frame $C^*$, see Fig.\ \ref{fig:schematic}, via
\begin{equation}
    ^{\text{C}^*}\mathbf{m}^* = \frac{1}{^{\text{C}^*}Z^*}\begin{bmatrix}^{\text{C}^*}X^*&^{\text{C}^*}Y^*&^{\text{C}^*}Z^*\end{bmatrix}^\text{T},
\end{equation}
which means it is observed by the camera as
\begin{equation}
    ^{\text{C}^*}\mathbf{p}^* = \mathbf{K}^{\text{C}^*}\mathbf{m}^*,
\end{equation}
in pixel coordinates $^{\text{C}^*}\mathbf{p}^*=\begin{bmatrix}u^*&v^*&1\end{bmatrix}^\text{T}$, with the camera's intrinsic parameters $\mathbf{K}$. Should the camera move under rotation $\mathbf{R}$ and translation $\mathbf{t}$, the points in normalized coordinates will change according to a homography $\mathbf{H}$ such that~\cite{benhimane2006homography}
\begin{equation}
    \frac{^{\text{C}}Z}{^{\text{C}^*}Z^*}{^{\text{C}}\mathbf{m}} = \mathbf{H}^{\text{C}^*}\mathbf{m}^*
\end{equation}
In pixel coordinates this can be written as
\begin{equation}
    \frac{^{\text{C}}Z}{^{\text{C}^*}Z^*}{^{\text{C}}\mathbf{p}} = \mathbf{G}{^{\text{C}^*}\mathbf{p}}^*,
\end{equation}
with the projective homography $\mathbf{G}$, for which the following relation holds
\begin{equation}
    \mathbf{H} = \mathbf{K}^{-1}\mathbf{G}\mathbf{K}.
    \label{eq:h_norm}
\end{equation}

As shown in \cite{benhimane2006homography}, the task error $^\text{C}\mathbf{e}_{\text{t}'} = \begin{bmatrix}^\text{C}\mathbf{e}_v & ^\text{C}\mathbf{e}_\omega\end{bmatrix}^\text{T}$ that urges to minimize the distance between the desired projection of $^\text{W}\mathbf{X}$, $^{\text{C}^*}\mathbf{m}^*$, and the current one $^\text{C}\mathbf{m}$, can be obtained purely from the homography that relates those points in normalized coordinates via
\begin{equation}
    \begin{split}
        ^\text{C}\mathbf{e}_v & = (\mathbf{H} - \mathbf{I})^{\text{C}^*}\mathbf{m}^*\\
        \left[^\text{C}\mathbf{e}_\omega\right]_\times & = \mathbf{H} - \mathbf{H}^\text{T},
    \end{split}
    \label{eq:dc}
\end{equation}
where $\left[^\text{C}\mathbf{e}_\omega\right]_\times$ is the skew symmetric matrix of $^\text{C}\mathbf{e}_\omega$. The task error $^\text{C}\mathbf{e}_{\text{t}'}$ is described in body coordinates. It can be transferred to the world frame W through rotation, which is proportional to camera frame's instantaneous velocity
\begin{equation}
    \begin{bmatrix}^\text{W}\mathbf{R}_\text{C} & \mathbf{0} \\ \mathbf{0} & ^\text{W}\mathbf{R}_\text{C}\end{bmatrix}{^\text{C}\mathbf{e}_{\text{t}'}} = ^\text{W}\mathbf{e}_{\text{t}'} \sim \mathbf{J}_{i+1}\dot{\mathbf{q}}
    \label{eq:body}
\end{equation}
where $^\text{W}\mathbf{R}_\text{C}$ is the rotation of the camera frame with respect to the world frame, and $\mathbf{J}_{i+1}$ is the camera frame's Jacobian, including its rotational contributions. Only 4 DOF can be controlled at a time after imposing the RCM, which constraints 2 DOF. To capture this, we introduce operator $\mathbf{P}$ that projects the camera frame body velocity onto the remaining DOF. Together with \ (\ref{eq:body}), this yields
\begin{equation}
    \mathbf{P}_{a/b}{^\text{C}\mathbf{e}_{\text{t}'}} = ^\text{C}\mathbf{e}_{\text{t}_{a/b}} \sim \mathbf{P}_{a/b} \begin{bmatrix}^\text{C}\mathbf{R}_\text{W} & \mathbf{0} \\ \mathbf{0} & ^\text{C}\mathbf{R}_\text{W}\end{bmatrix}\mathbf{J}_{i+1}\dot{\mathbf{q}}.
    \label{eq:proj}
\end{equation}
The projection operator $\mathbf{P}_{a/b}$ can take different forms, such that the task error is mapped onto any of the decoupled remaining DOF via
\begin{equation}
    \mathbf{P}_a = \begin{bmatrix}
        \mathbf{I}_{3\times3} & \multicolumn{3}{c}{\mathbf{0}_{3\times3}} \\
        \mathbf{0}_{1\times3} & 0 & 0 & 1
    \end{bmatrix},
    \mathbf{P}_b =
    \begin{bmatrix}
        0 & 0 & 1 & \mathbf{0}_{1\times3} \\
        \multicolumn{3}{c}{\mathbf{0}_{3\times3}} & \mathbf{I}_{3\times3}
    \end{bmatrix}.
    \label{eq:projection}
\end{equation}
Therefore, $\mathbf{P}_a$ maps the task error $^\text{C}\mathbf{e}_{\text{t}'}$ to its translational parts and the rotation about the optical axis, and $\mathbf{P}_b$ maps it to its rotational part and the error along the optical axis. We identify the case sensitive contributions of (\ref{eq:proj}) as the task Jacobian from (\ref{eq:task_jac}) and the task error from (\ref{eq:pid}), which yields
\begin{equation}
    \mathbf{J}_\text{t} = \mathbf{P}_{a/b} \begin{bmatrix}^\text{C}\mathbf{R}_\text{W} & \mathbf{0} \\ \mathbf{0} & ^\text{C}\mathbf{R}_\text{W}\end{bmatrix}\mathbf{J}_\text{i+1},\,
    %\mathbf{J}_\text{t} = 
    %\begin{cases}
    %    \mathbf{J}_{\text{t}_a}=\mathbf{P}_{a} \begin{bmatrix}^\text{C}\mathbf{R}_\text{W} & \mathbf{0} \\ \mathbf{0} & ^\text{C}\mathbf{R}_\text{W}\end{bmatrix}\mathbf{J}_\text{i+1} \\
    %    \mathbf{J}_{\text{t}_b}=\mathbf{P}_{b} \begin{bmatrix}^\text{C}\mathbf{R}_\text{W} & \mathbf{0} \\ \mathbf{0} & ^\text{C}\mathbf{R}_\text{W}\end{bmatrix}\mathbf{J}_\text{i+1}
    %\end{cases}
    \mathbf{e}^\text{p}_\text{t} = 
    \begin{cases}
        ^\text{C}\mathbf{e}_{\text{t}_a}=\begin{bmatrix}^\text{C}\mathbf{e}_v & ^\text{C}e_{\omega_z} \end{bmatrix}^T \\
        ^\text{C}\mathbf{e}_{\text{t}_b}=\begin{bmatrix}^\text{C}e_{v_z} & ^\text{C}\mathbf{e}_{\omega} \end{bmatrix}^T
    \end{cases}
    %\mathbf{J}_{\text{t}_{a/b}} = \mathbf{P}_{a/b} \begin{bmatrix}^\text{C}\mathbf{R}_\text{W} & \mathbf{0} \\ \mathbf{0} & ^\text{C}\mathbf{R}_\text{W}\end{bmatrix}\mathbf{J}_\text{i+1}, ^\text{C}\mathbf{e}_{\text{t}_{a}} = \begin{bmatrix}\mathbf{e}_v \\ e_{\omega_z} \end{bmatrix}, ^\text{C}\mathbf{e}_{\text{t}_{b}} = \begin{bmatrix}e_{v_z} \\ \mathbf{e}_{\omega} \end{bmatrix}
\end{equation}
This results in a task dimension $n_\text{t} = 4$, which means that together with the RCM objective that introduces 3 constraints and adds the additional DOF $\lambda$, the robot has to have at least 6 DOF.

\subsection{Processing Pipeline}
\label{sec:pipe}

An overview of the processing pipeline is depicted in Fig.\ \ref{fig:pipe}. A surgeon first controls the endoscope from within the camera's reference frame via the keyboard. Images of desired views are manually taken along the way and are used to construct a graph, wherein each vertex is an image. This is done within the \textit{homography generation} node. 

Initially, camera calibration considering an underlying radial/tangential distortion model is carried out to obtain the distortion coefficients and the camera intrinsics. Following that, an eye in hand calibration is performed to locate the camera frame position $^\text{W}\mathbf{x}_{i+1}$, and $^\text{W}\mathbf{x}_i$ is set to lie along the negative optical axis at the endoscope's length, see Fig.~\ref{fig:schematic}. 

Each image $\mathcal{I}$ that is processed within the \textit{homography generation} node undergoes distortion removal, followed by an intensity-based automatic detection of the endoscopic boundary circle. Therein, the image is smoothed with a bilateral filter and thresholded in HSV image space to obtain a binary mask. The circle's center is computed as the center of mass, and its radius is obtained from the steepest gradient of the marginalized binary mask. If the illumination in the endoscopic view is below a certain value, then the last known center and radius are considered instead. The maximum rectangle of a given aspect ratio that fits into the extracted circle is then cropped from the image $\mathcal{I}$. The crop is further rescaled. The camera intrinsics are updated accordingly from $\mathbf{K}$ to $\mathbf{K}^\prime$ by offsetting and scaling the principal point. 

Once the graph is built, the surgeon can browse through the image gallery, as shown in Fig.\ \ref{fig:pipe}, where each image corresponds to a vertex within the graph. The surgeon may then select a desired view and execute the visual servo. This will trigger a Dijkstra search for the closest path from the current vertex to the desired view/vertex at constant cost per edge. This path is executed sequentially. Therefore, the homography $\mathbf{G}$ from the next vertex to the current view is computed for the visual servo. To compute the homography, we extract image features and their descriptors with a SURF feature detector \cite{bay2006surf}. For each feature in the target view, the two nearest neighbors are found in the current view, and, via Lowe's ratio test \cite{lowe2004distinctive}, only features with distinctive descriptors are kept. The homography that maps features from the target view to the current view is then determined under RANSAC outlier rejection.

The updated camera intrinsics $\mathbf{K}^\prime$, together with the desired homography $\mathbf{G}$, are then sent down the pipeline to first transform the homography from pixel coordinates to normalized coordinates via (\ref{eq:h_norm}) and then to compute the desired task $^\text{C}\mathbf{e}_{\text{t}'}$ from (\ref{eq:dc}). The update rate of these operations are restricted by the camera frame rate, which is why the desired trocar position $^\text{W}\mathbf{x}_\text{trocar}$ is sent separately to the synchronizer node, see Fig.\ \ref{fig:pipe}. The synchronizer node takes a homography RCM visual servo action client, \textit{HRCMVSActionClient}, which request the \textit{HRCMVSActionServer} to execute the desired task $^\text{C}\mathbf{e}_{\text{t}'}$, while maintaining a desired trocar position $^\text{W}\mathbf{x}_\text{trocar}$. 

The \textit{HRCMVSActionServer} implements a state machine, which rejects infeasible requests. It computes the forward kinematics as well as the Jacobians and computes a joint position update $\Delta\mathbf{q}=\Delta t\dot{\mathbf{q}}$ via (\ref{eq:pid}) in the RCM implementation \textit{RCMImpl}, where $\Delta t$ is the control interval. The desired joint positions are then sent to the robot.

\section{EXPERIMENTAL SETUP}
\label{sec:experimental_setup}

This section gives an overview of the robotic system and its components in Sec.\ \ref{sec:robotic_system}. Following that, clinically relevant questions and the evaluation protocol are addressed in Sec.\ \ref{sec:clin_protocol}.

\subsection{Robotic System}
\label{sec:robotic_system}

Our experimental setup, see Fig.\ \ref{fig:setup}, uses a KUKA LBR Med 7 R800 robot. To control it, we created a bridge to ROS by wrapping the Fast Robot Interface (FRI) \cite{schreiber2010fast} with ROS' Hardware Interface functionality. We use a Storz Endocameleon Hopkins Telescope, from which we capture images using a Storz TH~102~H3-Z~FI camera head. The endoscope is mounted to the LBR Med 7 R800 robot with a custom designed 3D print. For illumination, we connect a Storz TL 300 Power LED 300 light source to the endoscope. The image feed is output to SDI, which we convert to HDMI with a Monoprice 3G SDI to HDMI converter. We then grab the HDMI signal with a DeckLink 4K Extreme 12G and stream it onto the ROS network.

% The acquired images are sent to the Storz TC 300 Image1 S H3-Link, which links the camera via DisplayPort to the Storz TC 200 Image1 S Connect. From there on, the image feed...

\subsection{Clinical Scenario Evaluation Protocol}
\label{sec:clin_protocol}

The proposed method is evaluated in the laparoscopic setup shown in Fig.\ \ref{fig:setup}. We utilize a Szabo Pelvic Trainer to simulate a trocar. A Kyoto Kagaku colon rectum tube is inserted into the Szabo Pelvic Trainer to model a laparoscopic view of the abdomen. The clinical procedure is then modeled as follows. The robot initially drives the endoscope to the trocar and $\lambda$ in (\ref{eq:lambda}) is set to $1$. Following that, the user mounts the camera and the light source to the laparoscope. The user then drives the laparoscope through the trocar into the phantom.

In the phantom, we identify four clinically relevant views of the scene. These views include an overview of the scene, a view of the tool insertion area towards the abdominal wall, and two close-ups, one for further examination. For visual servoing between these views in a clinical scenario, these three objectives are of importance

\begin{itemize}
    \item Servoing from any current to any target view.
    \item Servoing to target views under tool motion.
    \item Servoing to target views after phantom repositioning.
\end{itemize}

To address these scenarios, we design three experiments. For all experiments, after the laparoscope insertion, the user moves to the overview of the surgical scene, where the first image is taken through the GUI, which corresponds to the graph's root view/vertex, see Fig.\ \ref{fig:pipe}. We measure the deviation of the RCM from the trocar position, record the Mean Pairwise Distance (MPD) of SURF features from the current to the desired view, the task error, execution time, joint angles, and the camera tip position.

\subsubsection{Servoing from any current view to any target view}
\label{sec:clin_protocol_any}
In this scenario we investigate the system's capability to autonomously execute extreme view changes. The user moves from the overview to a close-up, from where the scene is further examined. The laparoscope is then moved manually to grant view of the tool insertion area. At this stage, tools would be inserted into the patient and the user would begin to operate. Therefore, the user selects the close-up view through the GUI and executes the autonomous visual servo towards it.

\subsubsection{Servoing to target views under tool motion}
\label{sec:clin_protocol_tool}
In this scenario we investigate autonomous visual servoing towards desired views under tool motion. Therefore, the user moves the laparoscope from the overview to the tool insertion area. Tools are then inserted and the user is asked to perform a sample task, which involves moving small LASTT Training Package rings. The visual servo simultaneously navigates back towards the overview.

\subsubsection{Servoing to target views after phantom repositioning}
\label{sec:clin_protocol_re}
In this scenario we investigate the system's invariance under patient motion. Therefore, we reposition the phantom and execute the visual servo to autonomously readjust the overview. We include both phantom rotation and tilting.

\section{RESULTS}
\label{sec:results}
In this section, we first present generic findings in Sec.\ \ref{sec:generic_res}, followed by quantitative measurements for the evaluation protocol from Sec.\ \ref{sec:clin_protocol}, in Sec.\ \ref{sec:clin_res}.
\begin{figure}
    \centering
    \begin{subfigure}
        \centering
        \includegraphics[width=\linewidth]{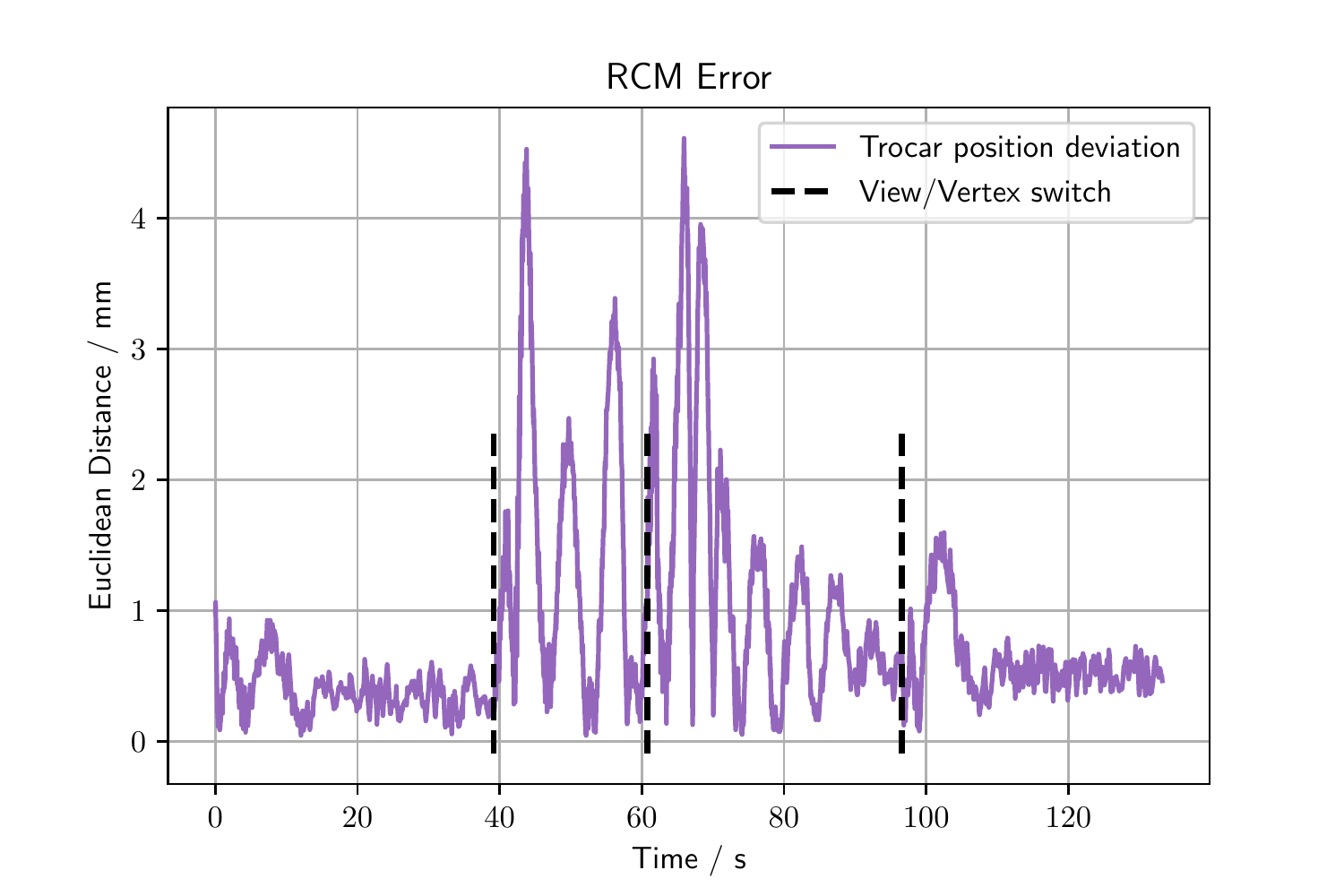}
    \end{subfigure}
    \begin{subfigure}
        \centering
        \includegraphics[width=\linewidth]{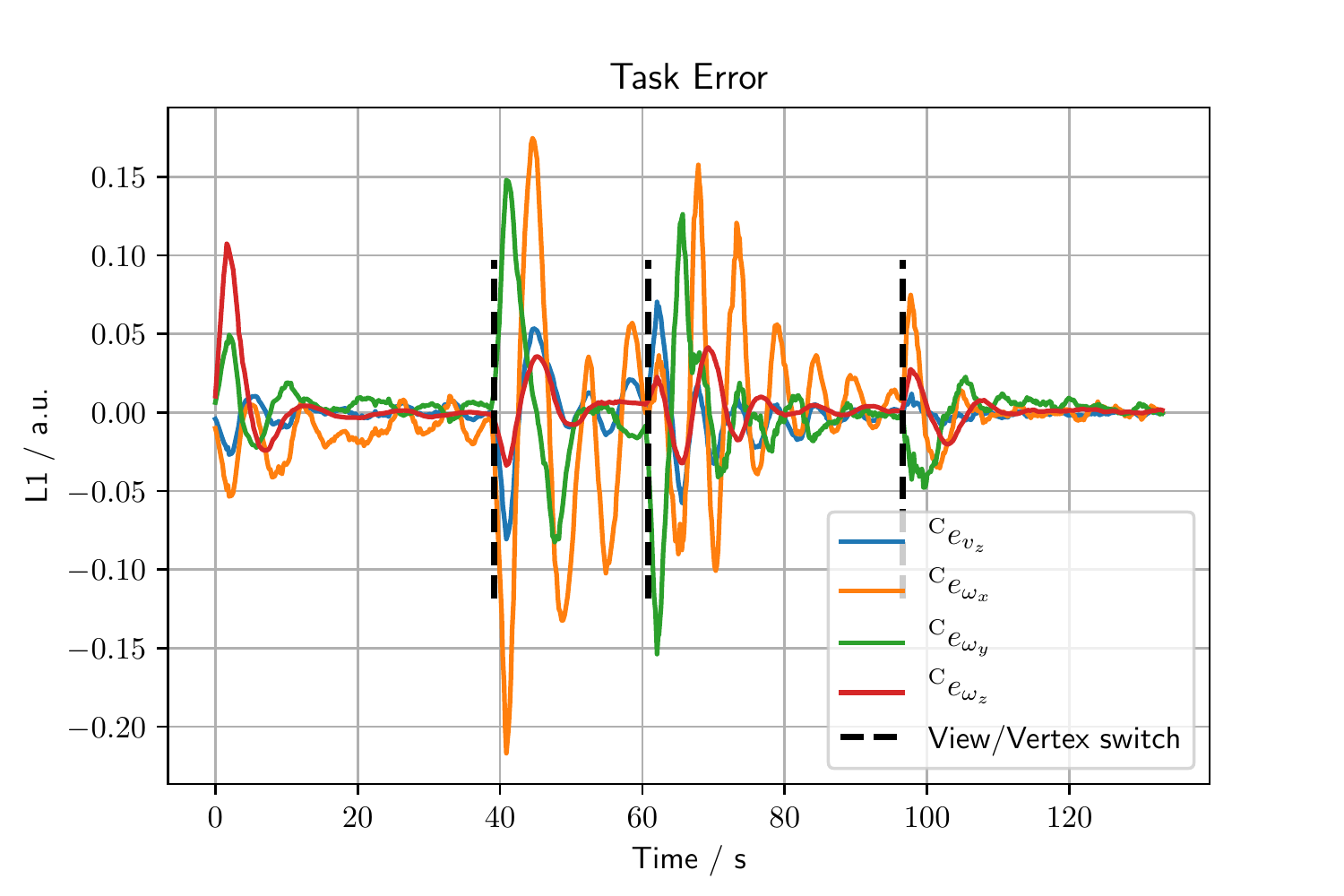}
    \end{subfigure}
    %\begin{subfigure}{.5\textwidth}
    %\centering
    %\includegraphics[width=.9\linewidth]{fig/visual_error.pdf}
    %\end{subfigure}
    \caption{RCM deviation (top) and task error evolution (bottom) over time for the protocol in Sec.\ \ref{sec:clin_protocol_any}. The visual servo autonomously servos from the tool insertion area to the close-up. Target views/vertices are updated along the way, as indicated by the black dotted lines.}
    \label{fig:errors}
\end{figure}
\subsection{Generic Results}
\label{sec:generic_res}
In practice we found that controlling the camera frame's rotational DOF, using $\mathbf{P}_b$ in (\ref{eq:projection}), leads to more stable solutions. We tried to invert the task part of the composite Jacobian from (\ref{eq:pid}) within the Nullspace of the RCM Jacobian, but obtained more flexible solutions by computing the pseudo-inverse as a damped least squares solution from the SVD with a damping factor of $5\mathrm{e}{-4}$. Empirically, we got good results with the following gain matrices
\begin{equation*}
    \begin{split}
        \mathbf{K}^\text{p} &= \diag(1.2, 1.5, 1.5, 1.8, 1\mathrm{e}{2}, 1\mathrm{e}{2}, 1\mathrm{e}{2}) \\
        \mathbf{K}^\text{i} &= \diag(3\mathrm{e}{-3}, 2.5\mathrm{e}{-3},2.5\mathrm{e}{-3}, 1.5\mathrm{e}{-3}, 0, 
    0, 0)\\
        \mathbf{K}^\text{d} &= \diag(6\mathrm{e}{-2}, 5\mathrm{e}{-2}, 5\mathrm{e}{-2}, 3\mathrm{e}{-2}, 0, 0, 0).
    \end{split}
\end{equation*}
The integral term therein helped remove a steady state error in the homography-based image alignments. The desired homography extraction proved noisy but correct on average, so we introduced a moving average filter on the task error $^\text{C}\mathbf{e}_\text{t}$ with a buffer length of $10$ at a frame rate of $30\,\text{fps}$. The sequential execution of desired views was greatly sped up by calling early convergence for intermediate vertices/views at a MPD of $5\,\text{pixels}$ and a final convergence at a MPD of $1.5\,\text{pixels}$. 

\subsection{Clinical Scenario Results}
\label{sec:clin_res}
\subsubsection{Servoing from any current view to any target view}
\label{sec:clin_res_any}

In this section we investigate the trajectory from tool insertion view to close-up, see Sec.\ \ref{sec:clin_protocol_any}. The task error and the RCM deviation from the trocar position are depicted in Fig.\ \ref{fig:errors}. It can be seen that the deviation from the trocar position stays below $4.6\,\text{mm}$, at an average deviation of $0.8\pm0.8\,\text{mm}$. The task error converges for all vertices/views. The final task error corresponds to a camera tip deviation of $0.4\,\text{mm}$, when compared to the desired position. The joint angles deviate on average by $8.2\pm6.0^\circ$ from the initial configuration.

\subsubsection{Servoing to target views under tool motion}
\label{sec:clin_res_tool}
\begin{figure}
    \centering
    \includegraphics[width=\linewidth]{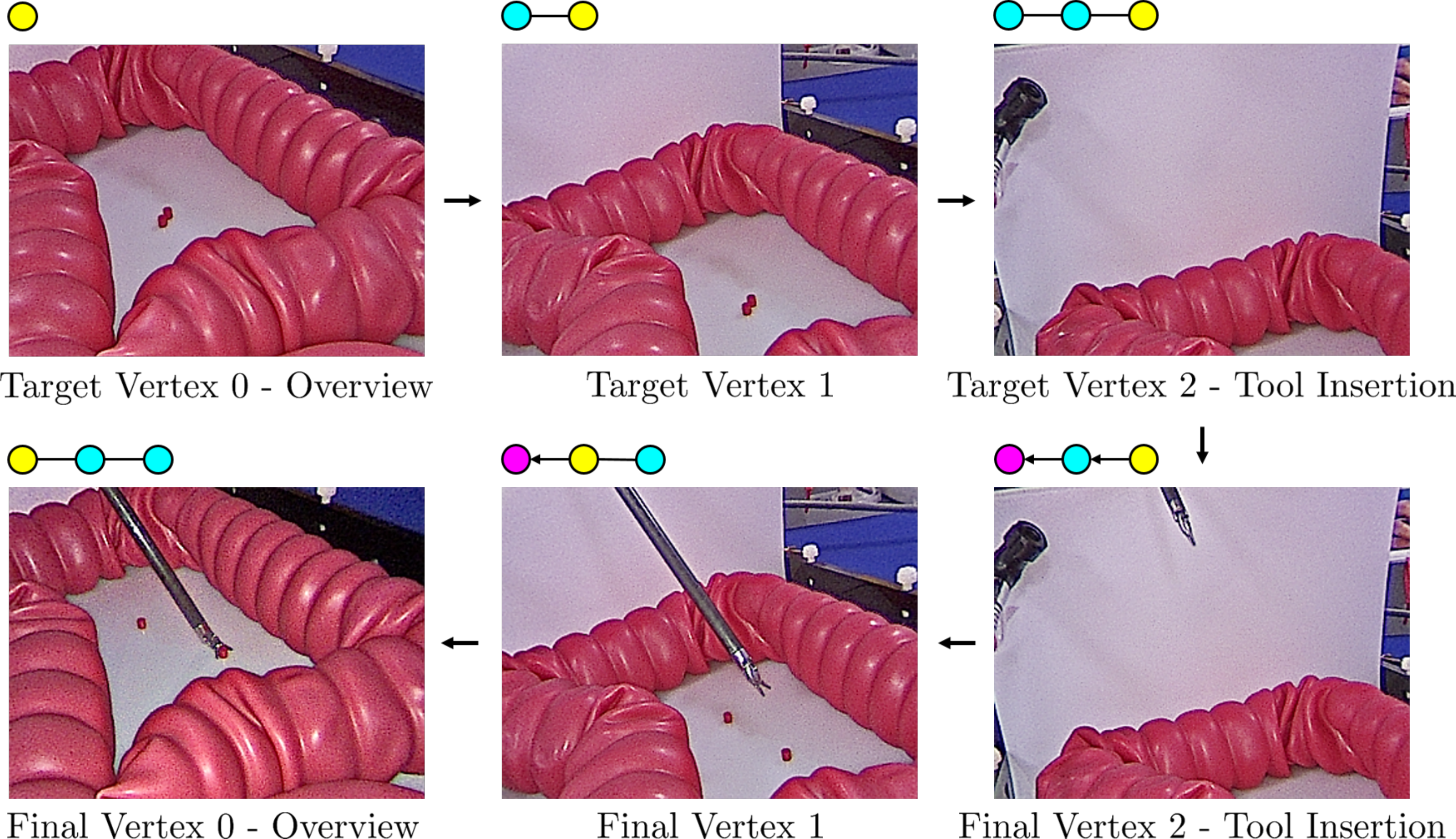}
    \caption{Servoing under tool motion, see Sec.\ \ref{sec:clin_protocol_tool}. Initially, the graph is built in manual control mode (top row), yellow indicates the current vertex. The visual servo is then executed to navigate back from the tool insertion to the overview (bottom row). Pink indicates the target vertex.}
    \label{fig:tool_insertion_trajectory}
\end{figure}

For this measurement, the visual servo navigates from the tool insertion area to the overview under tool motion, see Sec.\ \ref{sec:clin_protocol_tool}. The trajectory with all intermediate and the final vertex/view is shown in Fig.\ \ref{fig:tool_insertion_trajectory}. It can be seen that, despite tool motion, the visual servo converges at pixel accuracy towards the desired views. The final camera position deviates by $1.4\,\text{mm}$ to the desired one. The robot joint angles deviate on average by $1.1\pm1.1^\circ$ from the initial configuration. A video of this experiment is provided under footnote\ \ref{foot:vid}.

\subsubsection{Servoing to target views after phantom repositioning}
\label{sec:clin_res_re}
In this section we investigate the convergence of the visual error after phantom repositioning, see Sec.\ \ref{sec:clin_protocol_re}. We perform clockwise and counterclockwise repositioning as well as phantom tilting. We keep the trocar at the initial position. The camera frame then rotates and translates towards a position that minimizes the visual error. The translation $\Delta\mathbf{x}_{i+1}$ and the angle axis rotation angle $\alpha$ are listed in Tab.\ \ref{tab:repositioning}. It can be seen that the robotic laparoscope performs significant motion to readjust the view. The MPD is minimized to pixel range and the final deviation from the trocar remains in the submillimeter scale for all cases. 

\begin{table}
    \centering
    \caption{Clockwise (CW), and counterclockwise (CCW) repositioning, and phantom tilting, corresponding to the protocol in Sec.\ \ref{sec:clin_protocol_re}. $\Delta\mathbf{x}_{i+1}$ indicates the camera motion, $\alpha$ the angle axis rotation angle from initial to final camera rotation, $\Delta\mathbf{q}$ the joint angle position change, $\mathbf{e}_\text{RCM}$ the final deviation of the RCM from the trocar, and MPD the final visual error.}
    \begin{tabular}{c|ccc}
    
         Metric & CW & CCW & Tilt \\
         \hline
         $\Delta\mathbf{x}_{i+1}\,/\,\text{mm}$ & $10.4$ & $6.7$ & $4.7$ \\
         \hline
         $\alpha\,/\,^\circ$ & $16.6$ & $10.2$ & $4.8$ \\
         \hline
         $\Delta\mathbf{q}\,/\,^\circ$ & $20.5\pm12.0$ & $17.4\pm13.4$ & $2.6 \pm 2.3$ \\
         \hline
         $^\text{W}\mathbf{e}^\text{p}_\text{RCM}\,/\,\text{mm}$& $0.1$ & $0.2$ & $0.07$ \\
         \hline
         MPD / pixel & $3.2\pm2.5$ & $2.0\pm1.0$ & $1.4\pm1.2$
    \end{tabular}
    \label{tab:repositioning}
\end{table}
%\begin{table}[]
%    \centering
%    \caption{Clockwise (CW), counterclockwise (CCW), and tilt patient repositioning. $\Delta\mathbf{x}_{i+1}$ indicates the camera motion, $\Delta\mathbf{q}$ the joint angle position change, $\mathbf{e}_\text{RCM}$ the final deviation of the RCM from the trocar, and MPD the final visual error.}
%    \begin{tabular}{l|c|c|c|c|c}
%    
%         Type & $\Delta\mathbf{x}_{i+1}\,/\,\text{mm}$ & $\alpha\,/\,^\circ$ & $\Delta\mathbf{q}\,/\,^\circ$ & $\mathbf{e}_\text{RCM}\,/\,\text{mm}$ & MPD / pixel \\
%         \hline
%         CW   & $10.4$ & $16.6$ & $20.5\pm12.0$ & $0.1$  & $3.2\pm2.5$ \\
%         CCW  & $6.7$ & $10.2$ & $17.4\pm13.4$ & $0.2$  & $2.0\pm1.0$ \\
%         Tilt & $4.7$ & $4.8$ & $2.6 \pm 2.3$ & $0.07$ & $1.4\pm1.2$
%    \end{tabular}
%    \label{tab:repositioning}
%\end{table}

\section{CONCLUSION}
\label{sec:conclusions}
% summary
In this work we introduced a visual servo that is independent of depth information and explicit tool and camera positions. The introduced method simultaneously respects a programmable RCM. Our method was successfully integrated into a robotic setup and clinically relevant scenarios were investigated on an abdominal phantom.

% results discussion
It was shown in Sec.\ \ref{sec:clin_res_any} that the proposed composite Jacobian PID controller with homography-based task simultaneously minimizes the RCM and the visual servo objective. The integral term proved helpful to remove a steady state error in the image alignment. The homography estimation was noisy due to feature sparseness and required for average filtering. The graph representation allowed for visual servoing between images that were not relatable by a single homography transformation. In Sec.\ \ref{sec:clin_res_tool}, tools were successfully introduced into the scene. It is to be noted that the tools were initially not present in the target views, which removed potential image misalignment. In Sec.\ \ref{sec:clin_res_re} the phantom was repositioned significantly with a constant trocar position and image readjustment was successfully demonstrated. The MPD got close to perfect alignment, however, the trocar was possibly moved slightly during repositioning, which made perfect convergence not possible. The robot's joint angles did not always return to their initial configuration. The camera position converged in submillimeter range to its target.

% future work
We successfully demonstrated that our visual servo navigates the camera in submillimeter range without depth information or explicit tool and camera positions. This proves the future potential for safe patient application and it circumvents time-consuming registration procedures. As our setup has one redundant DOF, the robot did not always return to its initial configuration. This might be handled by introducing joint state objectives to the Jacobian's nullspace. While our visual servo is independent of registration procedures, the RCM requires initialization, and tracking. In future work, the controller might be updated as to incorporate force-torque sensing to update the RCM. Although the environment was mostly static, the homography estimation was noisy. In future research, one might, therefore, incorporate homography estimation that is invariant under object motion and robust under feature sparseness, using deep learning approaches, as shown in \cite{huber2021deep}.


%%%%%%%%%%%%%%%%%%%%%%%%%%%%%%%%%%%%%%%%%%%%%%%%%%%%%%%%%%%%%%%%%%%%%%%%%%%%%%%%

%%%%%%%%%%%%%%%%%%%%%%%%%%%%%%%%%%%%%%%%%%%%%%%%%%%%%%%%%%%%%%%%%%%%%%%%%%%%%%%%

%%%%%%%%%%%%%%%%%%%%%%%%%%%%%%%%%%%%%%%%%%%%%%%%%%%%%%%%%%%%%%%%%%%%%%%%%%%%%%%%
%\section*{APPENDIX}

\section*{ACKNOWLEDGMENT}
This work was supported by core and project funding from the Wellcome/EPSRC [WT203148/Z/16/Z; NS/A000049/1; WT101957; NS/A000027/1]. This project has received funding from the European Union's Horizon 2020 research and innovation programme under grant agreement No 101016985 (FAROS project). The authors gratefully acknowledge the support of Dr Carlo Seneci, Maleeha Al-Hamadani, Dr Chayanin Tangwiriyasakul, Dr Hongbing Liu, and Julius Bernth in the research that led to this manuscript. 

%%%%%%%%%%%%%%%%%%%%%%%%%%%%%%%%%%%%%%%%%%%%%%%%%%%%%%%%%%%%%%%%%%%%%%%%%%%%%%%%
\bibliographystyle{IEEEtran}
\bibliography{./IEEEtranBST/IEEEabrv, ./IEEEtranBST/IEEEmybib}
\end{document}